\documentclass[letter]{aa}  

\usepackage{graphicx}
\usepackage{txfonts}
\usepackage{hyperref}
\usepackage{multirow}
\hypersetup{
   colorlinks = true,
   citecolor ={cyan}
}
\begin{document}

   \title{Nitrogen rises to the top: evidence of enhanced mixing in very massive stars}


   \titlerunning{Nitrogen rises to the top}

   \author{Pablo Marchant
          \inst{1}
          \and
          Tomer Shenar\inst{2}
          }

   \institute{Department of Physics and Astronomy, Proeftuinstraat 86, N3, B-9000 Ghent, Belgium\\
              \email{pablo.marchant@ugent.be}
         \and
             The School of Physics and Astronomy, Tel Aviv University, Tel Aviv 6997801, Israel\\
             \email{tshenar@tauex.tau.ac.il}
             }

   \date{}

  \abstract{ Recent observations of young galaxies in the high-redshift Universe reveal signs of early enrichment of nitrogen. Extremely massive stars ($M \gtrsim 10^{2}-10^3\,M_\odot$) with strong stellar winds have been proposed as a potential driver of this phenomenon. Here, we show that the observed fraction of nitrogen rich stars with masses $\gtrsim 100M_\odot$ cannot be explained solely by mass loss, requiring significantly more efficient mixing beyond their convective cores than accounted in present evolutionary models. We compile a representative sample of 122 stars in the Tarantula Nebula of the Large Magellanic Cloud (LMC) with masses \mbox{$M \gtrsim  30\,M_\odot$}. Nearly all stars with masses $M \gtrsim 100\,M_\odot$ exhibit strong nitrogen enrichment, by factors $\gtrsim 5-10$. We demonstrate that this trend cannot be reproduced by varying assumptions on binary fraction, star formation history, or mass-loss rates within ranges predicted by current empirical and theoretical models. In contrast, enhanced core overshooting of $\alpha_{\rm ov}\gtrsim 1$ can  account for the observed enrichment, but leads to quasi-chemically homogeneous evolution that is inconsistent with the observed Hertzsprung–Russell diagram. While the origin of this discrepancy remains unclear, our results, in combination with observational and theoretical constraints on mass-loss rates, suggest the presence of efficient early mixing operating during or shortly after the formation of very massive stars. Such mixing models are currently not included in stellar evolution models. These findings have immediate implications for the formation, radial expansion, evolution, and final fates of stars at the upper mass end, and provide a potential pathway to explaining the rapid nitrogen enrichment observed in the high-redshift Universe.}


   \maketitle
%

\section{Introduction}

Despite the importance of massive stars in multiple areas of astrophysics, significant uncertainties remain in our understanding of their evolution \citep{Langer2012}. One important aspect that has received significant attention in the past decade is that most massive stars evolve in close binaries that undergo strong interactions \citep{Sana+2012, MarchantBodensteiner2024}. However, even in the absence of companions, important questions involving single-star physics remain, for example, concerning hydrodynamical processes both in their atmospheres and deep interior. One of these key uncertainties concerns the efficiency of mixing beyond the convective core. Enhanced mixing has been invoked to explain, for example, the properties of eclipsing binaries (e.g. \citealt{ClaretTorres2019, Higgins2019}), the width of the main sequence (e.g. \citealt{Fitzpatrick1990, Vink2010, Castro+2014, deBurgos2025, Lennon2026}), the  Humphreys-Davidson limit \citep{Gilkis+2021, Schootemeijer2026}, and the formation of Wolf-Rayet stars \citep{Maeder1987, Shenar+2020, Gilkis+2021, Schootemeijer+2024} and black-hole mergers \citep{MandeldeMink2016, Marchant+2016}. Beyond the main sequence, convective core overshooting in massive stars determines the size of the helium core, and thus, what type of compact object (and possible supernova) it produces \citep{Temaj+2024}.

In this paper, we make use of a sample of massive stars from the 30 Doradus region of the LMC to constrain the strength of mixing beyond the convective boundary at high masses, looking in particular at their surface nitrogen abundances. The large convective cores of stars in excess of $100M_\odot$, combined with their strong stellar winds, make it natural to expect them to be nitrogen enriched once they expose material that has been processed through the CNO cycle. For instance, \citet{Vink2023} has argued that very massive stars provide an explanation to observations of high redshift galaxies showing large nitrogen to oxygen ratios \citep{Senchyna+2024}. Here, we show that stellar winds are not sufficient to explain observed abundances of very massive stars, as the finite time needed to expose nitrogen enhanced layers through stellar winds would imply a significant fraction should have a normal nitrogen abundance at their surface, in contrast to observations. We argue that mixing beyond the convective core is significantly underestimated at high stellar masses.

\section{Sample}\label{sec:sample}

In this work, we focus on the massive-star population associated with  the Tarantula nebula  (\object{30 Dor})  in the sub-solar metallicity environment ($Z \approx 0.5\,Z_\odot$) of the Large Magellanic Cloud (LMC). 30 Dor, hosting thousands of massive stars \citep{Walborn1997, Evans+2011, Walborn2014} at a well-constrained distance ($d = 49.97\,$\,kpc \citealt{Pietrzynski2013}) and a  modest reddening \citep{Maiz2014}, arguably offers one of the best studied laboratories of massive stars at subsolar metallicity, including the most massive stars and binaries reported to date \citep{Crowther2010, Brands2022}. 

We compile a sample of 122 massive stars in the Tarantula region with available nitrogen abundances, combining several complementary spectroscopic studies  (Appendix\,\ref{app:sample}). The sample is primarily based on observations from the VLT-FLAMES Tarantula Survey \citep[VFTS;][]{Evans+2011, Bestenlehner2014}, supplemented by HST analyses of the dense R136 cluster \citep{Crowther2016, Bestenlehner+2020, Brands2022}, additional massive binaries from the Tarantula Massive Binary Monitoring survey \citep{Almeida2017, Mahy2020}, and a few well-studied systems not included in those surveys \citep{Tehrani2019, Shenar2017b, Shenar2021, Bestenlehner2022}. We further include a subset of nitrogen-rich Wolf–Rayet (WR) stars with properties consistent with possible main-sequence evolution, classified WNh and WN(h) stars \citep{deKoter1997}, while accounting for uncertainties in their evolutionary status. 
The sample is largely complete at high luminosities \citep[e.g.][]{Schneider2018Sci}, and constructed to avoid biases related to nitrogen enrichment or evolutionary selection; any residual uncertainties, primarily associated with the WR population, are tested and shown to not affect our conclusions.

\section{Analysis}\label{sec:analysis}

\begin{figure}
   \includegraphics[width=\columnwidth]{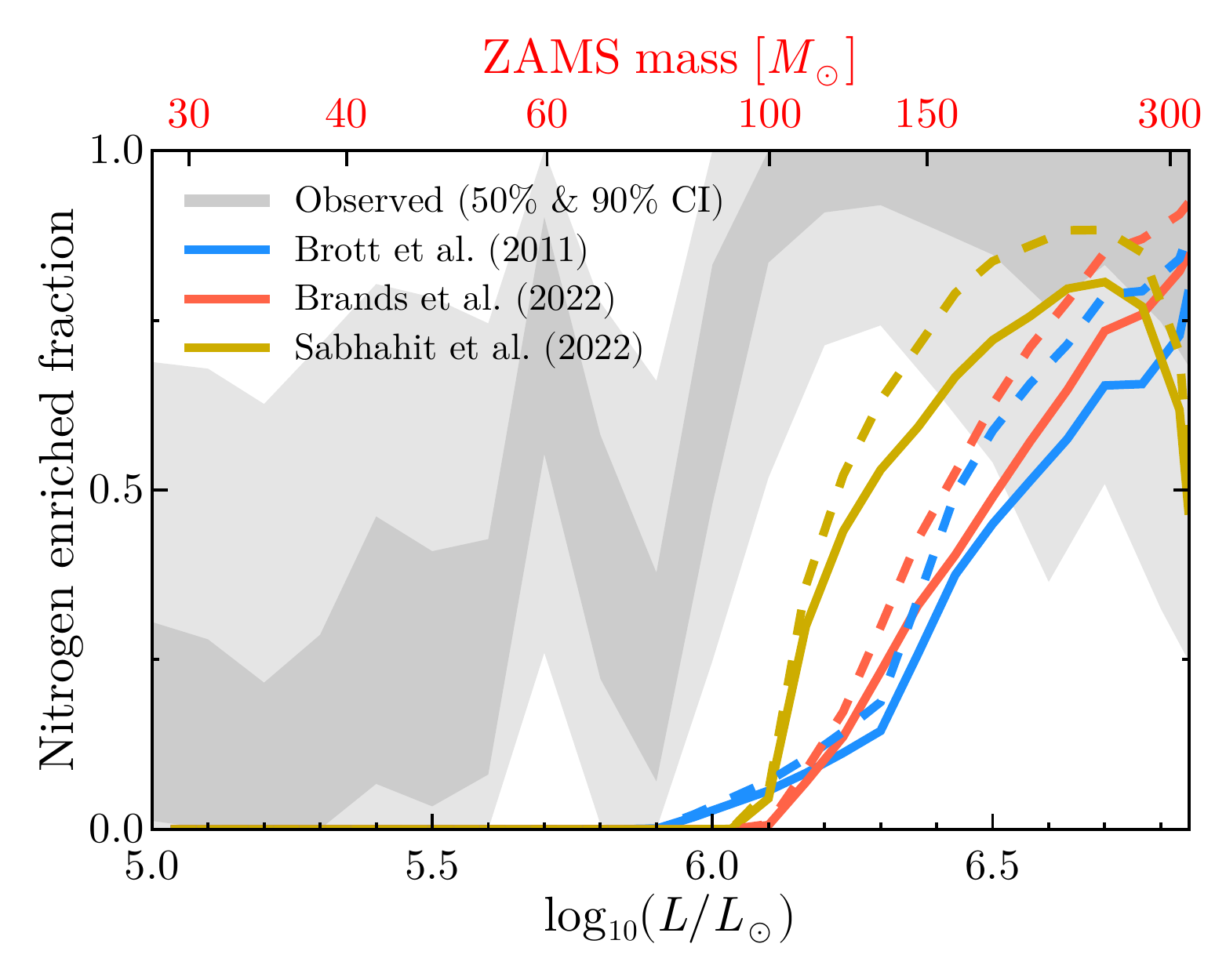}
   \caption{Observed fraction of nitrogen rich stars in 30 Doradus (gray band showing 50\% and 90\% highest density intervals) compared against theoretical predictions from single star models using different assumptions of wind physics with a value of overshoot of $\alpha_\text{ov}=0.335$ \citep{Brott+2011}. Solid lines indicate a constant star formation prior, while the dashed line lines uses a star formation history based on \citet{Schneider+2018_SFR}. Top axis indicates the mass of a ZAMS star with the corresponding luminosity.}\label{fig:enrichment}
\end{figure}

In order to compare our sample against theoretical models, we consider the fraction of enriched stars for a given luminosity, $P_\text{N-rich}(L)$. Ideally, given a sufficiently large number of stars and abundance constraints, it should be possible to compare population models directly against the distribution of nitrogen abundances versus luminosities. However, the scarcity of the most luminous stars, as well as the absence of specific abundance determinations for some (e.g. \citealt{Bestenlehner2014} provide only a qualitative estimate of nitrogen enrichment), requires this additional level of abstraction. In order to constrain $P_\text{N-rich}(L)$ we approximate it as a piecewise function and use a Bayesian analysis to determine it as well as to account for errors due to small samples at the highest luminosities (see Appendix \ref{app:Nrich_frac}). Figure \ref{fig:enrichment} shows that the fraction of enriched star is generally well below 50\% for luminosities lower than $10^6L_\odot$, with a strong shift towards a $100\%$ of stars being enriched at higher luminosities. The posterior distribution at those higher luminosities peaks at 100\% and the broadening uncertainty as luminosity increases is just a consequence of the low number of available sources.

From our evolutionary tracks coupled with an assumption on the star formation history of the 30 Doradus region we can directly predict the expected distribution of nitrogen enriched stars at a given luminosity (see Appendix \ref{app:physical_constraints}). Figure \ref{fig:enrichment} shows the theoretically predicted fractions of nitrogen enriched stars, considering the evolution of single non-rotating models with stellar winds prescribed according to either \citet{Brott+2011}, \citet{Brands2022} or \citet{Sabhahit+2022}. For the three cases overshooting is modelled using step overshooting with a dimensionless parameter $\alpha_\text{ov}=0.335$, as determined by \citet{Brott+2011}. All the models predict a negligible fraction of nitrogen rich stars below $10^6 L_\odot$, and although this fraction increases significantly at larger luminosities, it is still below our empirically determined values. Star formation history impacts results slightly; making use of the star formation history determined by \citet{Schneider+2018_SFR} for the LMC which peaks $2$\,Myrs ago leads, as expected, to an increase in the fraction of enriched stars compared to predictions done assuming a constant star formation history.

Our goal is to understand if enhanced mixing beyond the convective core can provide a better explanation to the large number of nitrogen rich stars in 30 Doradus at high luminosities. However, we cannot ignore the possible impact of binary interactions on the formation, which likely contributes to the presence of enriched stars at luminosities below $10^6L_\odot$. As a simplification, we assume that a fraction $f_\text{bin.\,prod.}$ of stars at any luminosity are nitrogen enriched binary products, while the remaining stars (which would correspond to both pre-interaction and effectively single stars) follow a distribution that matches our single star predictions. To account for different mixing efficiencies in the stellar envelope we consider different values of the ad-hoc overshooting parameter $\alpha_\mathrm{ov}$ which we take to be independent of mass and stellar age. In practice, overshooting could be both mass and age dependent \citep{Baraffe+2023, Johnston+2024}, or mixing could be driven by different processes such as rotation \citep{Maeder1987} or gravity waves excited at the boundary between the convective core and the radiative envelope \citep{Schatzman1993, RogersMcElwaine2017}. We use our models with fixed $\alpha_\text{ov}$ as a proxy on enhanced envelope mixing that can be improved in future studies with a physically motivated model.

\begin{figure}
   \includegraphics[width=\columnwidth]{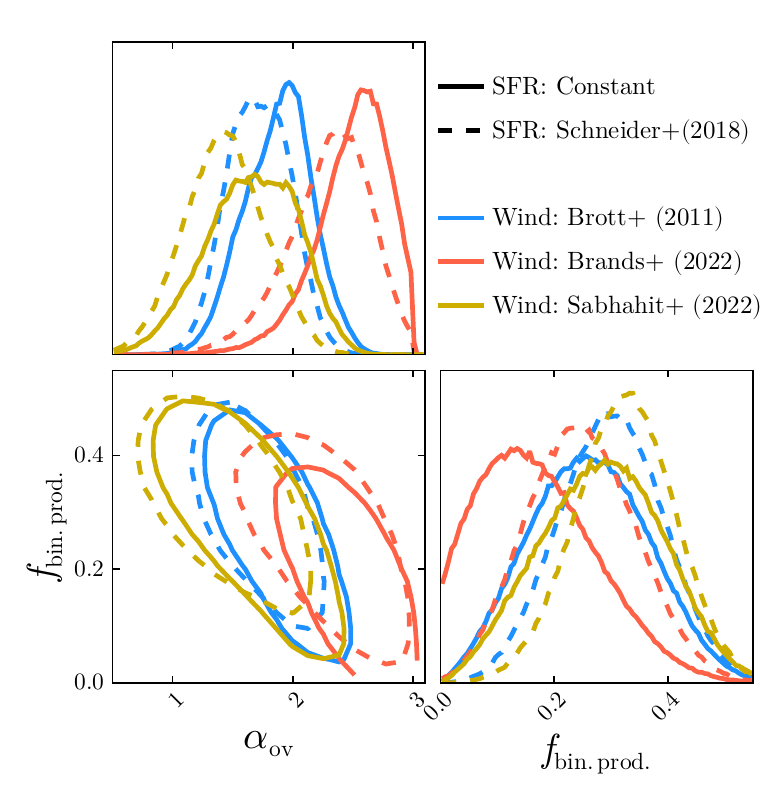}
   \caption{Posterior distributions for $\alpha_\text{ov}$ and $f_\text{bin.\,prod.}$ inferred by comparison to our 30 Doradus sample. Results for two star formation models (constant and following \citealt{Schneider+2018_SFR}) and three wind prescriptions \citep{Brott+2011,Brands2022,Sabhahit+2022} are shown. Panel for the 2D distribution of shows a 90\% highest density interval.}\label{fig:triangle}
\end{figure}

Following the method described in Appendix \ref{app:physical_constraints}, we perform a Bayesian analysis to constrain the possible values of $\alpha_\text{ov}$ and $f_\text{bin.\,prod.}$ accounting for the different models of star formation. Figure \ref{fig:triangle} shows a corner plot of the inferred posterior distributions, while Table \ref{tab:inferred} summarizes the inferred values in each case. Independent of the star formation history and wind model, $\alpha_\text{ov}$ and $f_\text{bin.\,prod.}$ exhibit an anti-correlation and $f_\text{bin.\,prod.}$ is restricted to values below $0.5$. Interestingly, the results for $\alpha_\text{ov}$ support a value in excess of unity, much larger than is normally adopted in stellar simulations of massive stars. This holds unless we consider mass loss rates in excess of current prescriptions (see Appendix \ref{app:mdot}).

\begingroup
\setlength{\tabcolsep}{8pt} 
\renewcommand{\arraystretch}{1.5} 
\begin{table}
\caption{Inferred 90\% highest density intervals for $\alpha_\text{ov}$ and $f_\text{bin.\,prod.}$ in our 30 Doradus sample considering different wind prescriptions and star formation histories.}              
\label{table:1}      
\centering                                      
\begin{tabular}{c | c c | c c}          
\hline\hline                        
 & \multicolumn{2}{|c|}{SFR: Constant} & \multicolumn{2}{|c}{SFR: Schneider} \\    
\hline
Winds & $\alpha_\text{ov}$ & $f_\text{bin.\,prod.}$ & $\alpha_\text{ov}$ & $f_\text{bin.\,prod.}$ \\    
\hline
Brott & $1.97^{+0.35}_{-0.56}$ & $0.26^{+0.17}_{-0.17}$ & $1.66^{+0.48}_{-0.38}$ &  $0.29^{+0.15}_{-0.15}$\\
Brands & $2.57^{+0.43}_{-0.51}$ & $0.14^{+0.18}_{-0.14}$ & $2.36^{+0.51}_{-0.59}$ & $0.24^{+0.16}_{-0.15}$ \\
Sabhahit & $1.68^{+0.59}_{-0.64}$ & $0.29^{+0.16}_{-0.18}$ & $1.42^{+0.56}_{-0.53}$ & $0.33^{+0.14}_{-0.16}$ \\
\hline                                             
\end{tabular}\label{tab:inferred}
\end{table}
\endgroup

\section{Discussion and Conclusions}\label{sec:discussion}
We have shown that the population of very massive stars in the 30 Doradus star forming region of the LMC indicates the presence of strong mixing processes operating beyond the boundaries of their convective cores.
In our model, we have considered overshooting as the source of this additional mixing, although other processes can be in operation and explain the large fraction of nitrogen enriched stars observed at high luminosities. Rotational mixing has been a commonly invoked mechanism to explain surface nitrogen enrichment, possibly leading to chemically homogeneous evolution for the fastest rotators \citep{Maeder1987, Brott+2011}. However, the observed rotation rates of O-type stars \citep[e.g.,][]{RamirezAgudelo+2015, Ramachandran+2019, Holgado+2022, Lennon2026} suggest that the vast majority of massive stars do not rotate fast enough to experience significant mixing. The physical process leading to efficient mixing could possibly operate independent of rotation and, if driven by the physics of the deep stellar interior where matter is fully ionized, also be metallicity independent. 

The impact of additional mixing in massive stars, capable of bringing the products of core hydrogen burning to its surface at an early stage, is very significant. Strong mixing leading to homogeneous evolution can result in the formation of gravitational wave sources from initially short period massive binaries \citep{MandeldeMink2016, Marchant+2016}. The production of larger helium cores can ease the production of Wolf-Rayet stars at low metallicities from self-stripping, possibly explaining the large fraction of single Wolf-Rayet stars in the Small Magellanic Cloud \citep{Shenar2019, Shenar+2020, Massey2023, Schootemeijer+2024}. The impact of strong core overshooting on the post main-sequence evolution of massive stars serves as a possible explanation of the Humphreys-Davidson limit \citep{Davies+2018, Gilkis+2021, Sabhahit+2021, Schootemeijer2026}, significantly restricting the expansion of very massive stars and the orbital separations at which they would experience Roche-lobe overflow \citep[e.g.][]{Romagnolo2023}.

.

Regardless of the mixing agent, the fact that models of massive stars significantly underpredict the amount of nitrogen in their atmospheres could have substantial implications for interpretation of the high redshift Universe.  Recent observations reveal strong nitrogen enrichment was reported in high redshift galaxies \citep{Topping+2024, Ji+2024, Cameron+2023, Bunker+2023, Senchyna+2024}. Nitrogen-rich massive stars therefore provide a natural local-Universe anchor for interpreting the prominent nitrogen signatures observed at high redshift, with important implications for stellar yields, feedback, and the calibration of nebular abundance diagnostics.

\begin{figure}
   \includegraphics[width=\columnwidth]{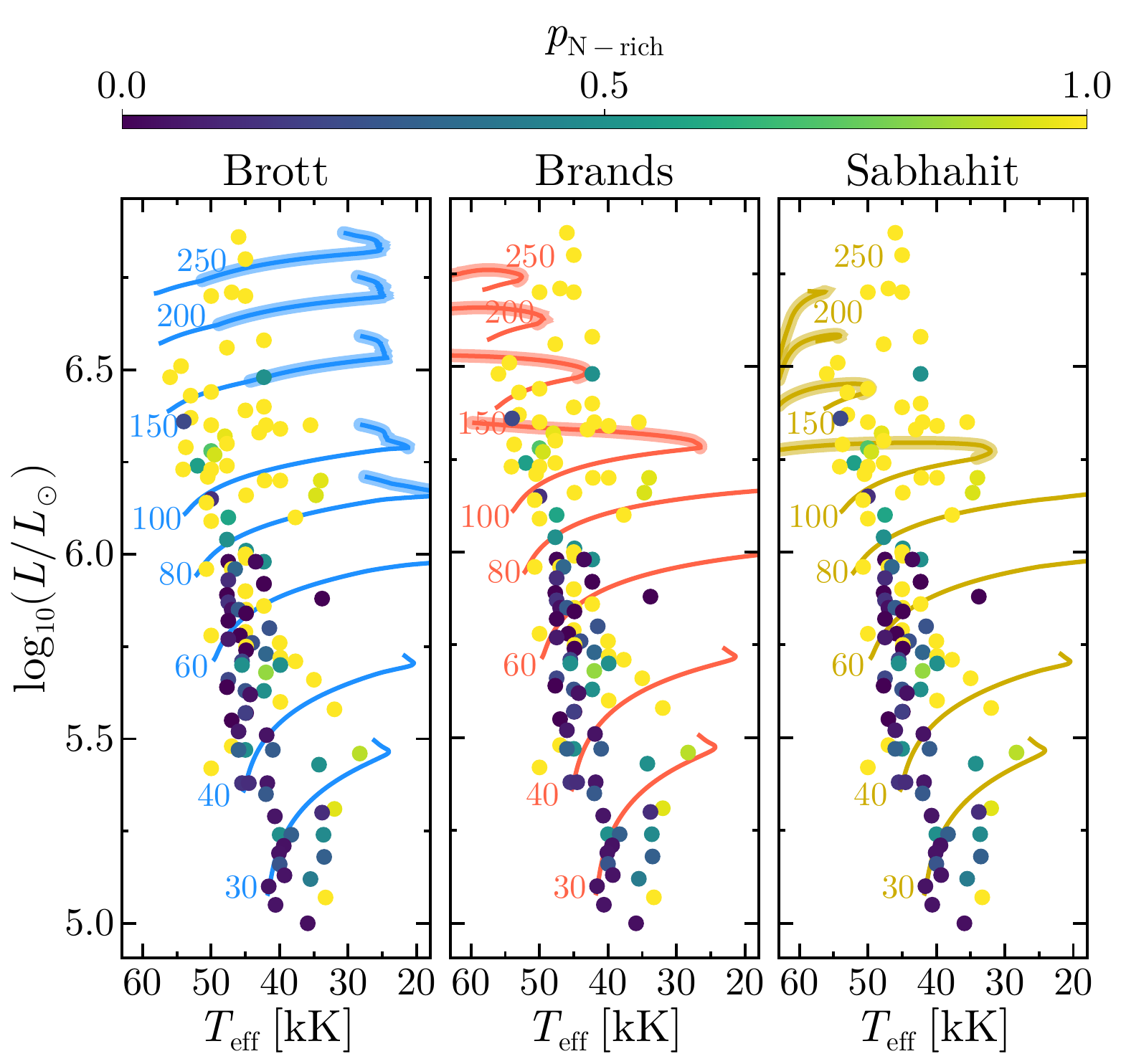}
   \caption{Location of stars in our sample indicating the probability that they are nitrogen enriched, plotted together with main-sequence evolutionary tracks with different assumptions on wind physics and a step overshoot parameter of $\alpha_\text{ov}=0.335$. Numbers indicate initial masses of the evolutionary tracks in $M_\odot$, while think lines indicate the phases where the nitrogen abundance exceeds the carbon baseline of the LMC.}\label{fig:HR_sample}
\end{figure}

\begin{figure}
   \includegraphics[width=\columnwidth]{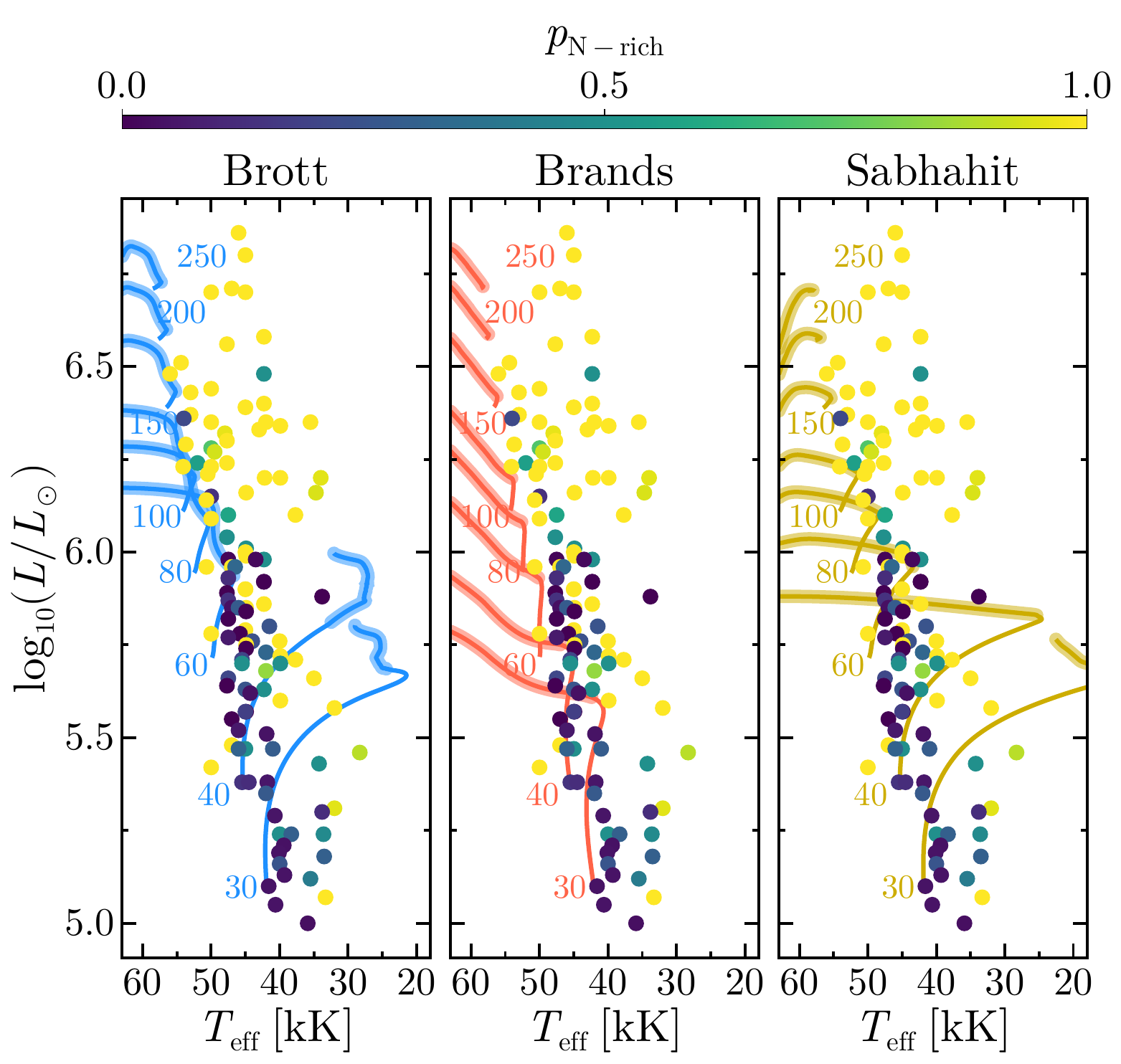}
   \caption{Same as Figure \ref{fig:HR_sample} but making use of the $\alpha_\text{ov}$ values given in Table \ref{tab:inferred} for the \citet{Schneider+2018_SFR} star formation history model ($\alpha_\text{ov}=$1.66, 2.36 and 1.42 respectively for each panel from left to right).}\label{fig:HR_sample_ov}
\end{figure}

Although we have shown that high overshooting can explain observed surface nitrogen abundances, such efficient mixing operating throughout the stellar lifetime also bears 
consequences which are in tension with observations. For example, Figs.\,\ref{fig:HR_sample}, \ref{fig:HR_sample_ov} show a Hertzsprung-Russell diagram (HRD) with MESA tracks computed with distinct mass-loss and overshooting prescriptions (see caption), compared to the measured positions of our sample. Evidently large overshooting causes stars in excess of $60M_\odot$ to evolve homogeneously, remaining hotter than the observed population of 30 Doradus at luminosities in excess of $10^6L_\odot$.

However, in principle, enhanced mixing needs not operate throughout the entire lifetime of the star. The conversion of carbon into nitrogen through the CN cycle occurs very early during the main sequence, so very efficient mixing restricted to an early stage could explain the observed enrichment without the larger consequences of long-term mixing. The  discrepancy seen in the HRD can thus be mitigated by a decreasing efficiency of mixing as a star evolves in the main sequence, or be partially explained by the effect of radiative hydrodynamical processes not accounted for in evolutionary models, which lower the effective temperatures in the atmospheres of stars near the Eddington limit \citep{Poniatowski+2021, Moens+2025}. An alternative explanation that does not involve enhanced mixing in stellar interiors would be that above a certain stellar mass most stars are the product of mergers through which nuclear burning products are exposed.

Nevertheless, the development of more physically motivated models for mixing beyond the convective boundary rather than ad-hoc parameterizations is of critical importance to describe the abundances of very massive stars as well as their final evolutionary products and their impact at large scales. 

\begin{acknowledgements}
      PM acknowledges support from the FWO senior fellowship number 12ZY523N. This work has received funding from the European Research Council (ERC) under the European Union’s Horizon 2020 research and innovation programme (101165213/Star-Grasp, 101164755/METAL). The authors also thank the anonymous referee for their helpful feedback.
\end{acknowledgements}


\begin{appendix} 

\section{Sample construction}
\label{app:sample}

We compiled a list of 122 massive stars in the Tarantula region for which nitrogen abundances were derived or estimated, listed by descending luminosities in Appendix\,\ref{app:sample} (Tables\,\ref{tab:Log1} and \ref{tab:Log2}).
The majority of targets were analysed by \citet{Bestenlehner2014, Bestenlehner2020} and \cite{Brands2022}. \citet{Bestenlehner2014} used data acquired with the Fibre Large Array Multi Element Spectrograph (FLAMES) multi-fiber spectrograph mounted the Very Large Telescpoe (VLT) in the framework of the VLT-FLAMES Tarantula Survey (VFTS, \citealt{Evans+2011}). However, these data excluded  the  R136 cluster, which is at the core of the Tarantula and which is too dense for the FLAMES instrument.  This sample is complemented by analyses of R136 by \citet{Bestenlehner+2020} and \citet{Brands2022}, who, respectively, used optical and UV data acquired with the Hubble Space Telescope (HST) by \citet{Crowther2016}. We also append the binary components of double-lined spectroscopic (SB2) binaries analysed by \citet{Mahy2020} using FLAMES data acquired in the framework of the Tarantula Massive Binary Monitoring survey (PI: Sana, \citealt{Almeida2017}). We add the few massive binaries Mk\,34 \citep{Tehrani2019}, R\,144 \citep{Shenar2017b}, R\,145 \citep{Shenar2021}, and Mk\,33Na \citep{Bestenlehner2022}. For binaries, the suffixes "-P" and "-S" denote the "primary" and "secondary" components, respectively. The separations between the binary components included here are large enough such that tidal forces can be neglected.

Finally, we append to this list a handful of late-type WNh or WN(h) stars in the Tarantula exhibiting significant or moderate hydrogen enhancement, respectively. Very massive stars can appear as WNh or WN(h) stars already on the main sequence \citep{deKoter1997}, and it is therefore important to consider this population as well. However, it is not always easy to tell apart main sequence WR stars from post main sequence (i.e., classical) WR stars (e.g., see the case of the binary R\,144, \citealt{Shenar2021}). To include as many targets as possible,  we included all WR stars in the Tarantula  with $T_{\rm eff} \le 60\,$kK and $X_{\rm H} > 0.1$. Likely, some of these targets are classical WR stars rather than main sequence ones (e.g., BAT99~89, with its relatively low hydrogen content and bolometric luminosity). To ensure that our results are robust against this uncertainty, we perform the analysis twice, where we exclude stars with hydrogen mass fractions smaller than 0.4 (see Appendix \ref{section:reanalysis}). 

For all objects, we provide their effective temperatures $T_{\rm eff}$, bolometric luminosities $\log L$, surface hydrogen mass fractions $X_{\rm H}$, and estimated ages. To simplify the analysis, we account only for a probability $p_\text{N-rich}$ of a source being nitrogen enriched given observational constraints (see Appendix \ref{app:Nrich_frac}). In particular, we define an object as nitrogen rich if its surface nitrogen abundance exceeds the baseline value of carbon in the LMC, indicating significant CNO processing. . While the compiled sample is likely largely complete for very luminous stars in the Tarantula ($\log L \gtrsim 5.5\,[L_\odot]$), it is not necessarily complete at lower luminosities. However, completeness is not essential in the context of our study. Important, instead, is that none of the samples we include in the compilation is biased in age or other properties that may correlate with the nitrogen abundance. For this reason, we do not include studies such as \citet{Dufton2018}, who focused on stars with low rotation, or \citet{Grin2017} and \citet{McEvoy2015}, who only studied evolved OB-type stars. The main biases in our sample may originate in the WR population, as discussed above. However, we show that our main conclusions do not depend on the inclusion of the WR sample.

All sources used in this work are listed in Tables \ref{tab:Log1} and \ref{tab:Log2}. In each case we assign a probability for the star to be enriched in nitrogen, by which we mean that its surface nitrogen abundance $12+\log{n_{N14}/n_{H1}}$ exceeds the carbon baseline of $7.75$, well in excess of the nitrogen baseline abundance of the LMC of $6.90$ (both adopted from the work of \citealt{Brott+2011}). For all sources that have a measurement of nitrogen abundance with an error, we estimate the probability of being enriched ($p_\text{N-rich}$) by considering a normal distribution for the abundance and taking the fraction of this distribution that exceeds $7.75$. 

For various sources there is an abundance estimate but no error on it, in which case we adopt an uncertainty of $0.5$\,dex. In multiple instances, particularly at high luminosities, no quantitative abundance constraint is available. This is the case for multiple sources in the work of \citet{Bestenlehner2014}, where stars are assigned flags 'e', 'ne' and 'n' depending on whether or not they exhibit nitrogen enhancement, nitrogen enhancement is uncertain, or if they have a normal nitrogen abundance for the LMC. For these cases we take enhancement probabilities of 1.0, 0.5 and 0.0 respectively. We note that the metallicity baseline used by \citet{Bestenlehner2014} is roughly a factor two larger than adopted here. However, this leads to a conservative estimate of the fraction of nitrogen-enhanced stars in our work: stars marked as enriched (e) in \citet{Bestenlehner2014} would have enrichment factors $\gtrsim 20$ in our models. Hence, our reported fractions of enriched stars represent lower limits when considering the sample of \citet{Bestenlehner2014}. From the entire sample three objects (R136a8, H31 and H47) have reported nitrogen abundances despite there being no reliable lines to determine it. In our analysis we exclude R136a8, but keep H31 and H47 to provide a pessimistic perspective for nitrogen enrichment, as these two sources lie very close to the transition in the nitrogen enriched fraction seen at $10^6L_\odot$.

\begin{table}
\centering
\caption{Log of all objects used in our analysis. Typical $1\sigma$ errors on $\log L$, $T_{\rm eff}$, $X_{\rm H}$, and ages are $0.1\,$dex and $2000\,$K, 0.05, and 0.50\,Myr, respectively. Probability of a star being nitrogen rich ($p_\text{N-rich}$) is determined as described in Appendix \ref{app:Nrich_frac}. Suffixes '-P' and '-S' denote respectively the primary and secondary components of binaries.}
\resizebox{.47\textwidth}{!}{\begin{tabular}{clcccccc}
\hline  \hline
Index & Object & $\log T_{\rm eff}$ & $\log L$ & $X_{\rm H}$ & Age & $p_\text{N-rich}$ & References \\ 
 &  & kK & $[L_\odot]$ &  & Myr &  &  \\ 
\hline
1 & R136a1 & 46000 & 6.86 & 0.50 & 1.14 &  1.00 & 1;2\\ 
2 & BAT99 102 & 45000 & 6.80 & 0.40 &  - &  1.00 & 9\\ 
3 & R136a2 & 47000 & 6.71 & 0.45 & 1.34 &  1.00 & 1;2\\ 
4 & BAT99 98 & 45000 & 6.70 & 0.60 &  - &  1.00 & 9\\ 
5 & R136a3 & 50000 & 6.70 & 0.45 & 1.28 &  1.00 & 1;2\\ 
6 & VFTS 1025 & 42300 & 6.58 & 0.30 & 1.80 &  1.00 & 3;4\\ 
7 & Mk 42 & 47700 & 6.56 & 0.68 &  - &  1.00 & 3;4\\ 
8 & VFTS 682 & 54400 & 6.51 & 0.45 & 1.00 &  1.00 & 3;4\\ 
9 & VFTS 1022 & 42300 & 6.48 & 0.75 & 1.10 &  0.50 & 3;4\\ 
10 & BAT99 112 & 56000 & 6.48 & 0.20 &  - &  1.00 & 9\\ 
11 & R144a & 50000 & 6.44 & 0.35 & 2.00 &  1.00 & 7\\ 
12 & Mk 34a & 53000 & 6.43 & 0.65 & 0.50 &  1.00 & 8\\ 
13 & VFTS 482 & 42300 & 6.40 & 0.68 & 1.20 &  1.00 & 3;4\\ 
14 & R144b & 45000 & 6.39 & 0.40 & 2.00 &  1.00 & 7\\ 
15 & Mk 34b & 53000 & 6.37 & 0.65 & 0.60 &  1.00 & 8\\ 
16 & R136a7\tablefootmark{a} & 54000 & 6.36 & 0.70 & 0.50 &  0.22 & 1;2\\ 
17 & BAT99 96 & 42000 & 6.35 & 0.20 &  - &  1.00 & 9\\ 
18 & R136b & 35500 & 6.35 & 0.70 & 2.02 &  1.00 & 1;2\\ 
19 & R145-P & 50000 & 6.35 & 0.40 & 2.30 &  1.00 & 6\\ 
20 & VFTS 1021 & 39900 & 6.34 & 0.75 & 2.10 &  1.00 & 3;4\\ 
21 & R145-S & 43000 & 6.33 & 0.50 & 2.30 &  1.00 & 6\\ 
22 & R136a5 & 48000 & 6.32 & 0.70 & 0.98 &  0.97 & 1;2\\ 
23 & VFTS 545 & 47700 & 6.30 & 0.75 & 1.00 &  1.00 & 3;4\\ 
24 & VFTS 617 & 53700 & 6.29 & 0.38 & 2.00 &  1.00 & 3;4\\ 
25 & R136a4 & 50000 & 6.28 & 0.74 & 0.84 &  0.73 & 1;2\\ 
26 & H36 & 49500 & 6.27 & 0.74 & 0.90 &  0.93 & 1;2\\ 
27 & VFTS 506 & 47700 & 6.24 & 0.75 &  - &  1.00 & 3;4\\ 
28 & R136a6 & 52000 & 6.24 & 0.74 & 0.76 &  0.56 & 1;2\\ 
29 & VFTS 16 & 54100 & 6.23 & 0.75 & 0.70 &  1.00 & 3;4;11\\ 
30 & BAT99 122 & 50000 & 6.23 & 0.20 &  - &  1.00 & 9\\ 
31 & VFTS 1017 & 50500 & 6.21 & 0.15 & 2.00 &  1.00 & 3;4\\ 
32 & VFTS 527-P & 34000 & 6.20 & 0.74 & 2.00 &  0.96 & 5\\ 
33 & VFTS 457 & 39900 & 6.20 & 0.60 & 2.00 &  1.00 & 3;4\\ 
34 & VFTS 1001 & 42200 & 6.20 & 0.75 & 3.00 &  1.00 & 3;4\\ 
35 & R136a8\tablefootmark{b} & 49500 & 6.17 & 0.74 & 0.90 &  0.07 & 1;2\\ 
36 & VFTS 542 & 44900 & 6.16 & 0.53 & 2.20 &  1.00 & 3;4\\ 
37 & VFTS 527-S & 34700 & 6.16 & 0.74 & 2.10 &  0.94 & 5\\ 
38 & Mk 33Na-P & 50000 & 6.15 & 0.74 & 0.90 &  0.14 & 12\\ 
39 & VFTS 621 & 50700 & 6.14 & 0.75 &  - &  1.00 & 3;4\\ 
40 & VFTS 259 & 37700 & 6.10 & 0.68 & 2.30 &  1.00 & 3;4\\ 
41 & H46 & 47500 & 6.10 & 0.74 & 1.30 &  0.59 & 1;2\\ 
42 & BAT99 103-P & 50000 & 6.09 & 0.20 &  - &  1.00 & 10\\ 
43 & VFTS 512 & 47700 & 6.04 & 0.75 &  - &  0.50 & 3;4\\ 
44 & VFTS 267 & 44900 & 6.01 & 0.75 & 1.60 &  0.50 & 3;4\\ 
45 & BAT99 68 & 45000 & 6.00 & 0.60 &  - &  1.00 & 9\\ 
46 & VFTS 599 & 44900 & 5.99 & 0.75 & 1.20 &  1.00 & 3;4\\ 
47 & VFTS 603-P & 42300 & 5.98 & 0.75 &  - &  0.50 & 3;4\\ 
48 & H31\tablefootmark{b} & 47500 & 5.98 & 0.74 & 1.26 &  0.01 & 1;2\\ 
49 & H47\tablefootmark{b} & 43500 & 5.98 & 0.74 & 1.70 &  0.02 & 1;2\\ 
50 & VFTS 72 & 50700 & 5.96 & 0.75 & 0.40 &  1.00 & 3;4\\ 
51 & BAT99 67 & 47000 & 5.96 & 0.30 &  - &  1.00 & 9\\ 
52 & H48 & 46500 & 5.96 & 0.74 & 1.40 &  0.33 & 1;2\\ 
\hline
\end{tabular}}
\tablefoot{References: (1) \citet{Brands2022}, (2) \citet{Bestenlehner2020}, (3) \citet{Bestenlehner2014}, (4) For ages: \citet{Schneider2018Sci}, (5) \citet{Mahy2020}, (6) \citet{Shenar2017b}, (7) \citet{Shenar2021}, (8) \citet{Tehrani2019}, (9) \citet{Hainich2014}, (10) \citet{Shenar2019}, (11) \citet{Grin2017}, (12) \citet{Bestenlehner2022}, (13)  \citet{Shenar2022}\\
\tablefoottext{a}{ Adopted from  \citet{Bestenlehner2020} due to large error by \citet{Brands2022} }\tablefoottext{b}{No reliable N-lines available}\tablefoottext{c}{Contradictory between \citet{Brands2022} and \citet{Bestenlehner2020}}\tablefoottext{d}{Contradictory between \citet{Brands2022} and \citet{Bestenlehner2020}}.}
\label{tab:Log1}
\end{table}
\begin{table}
\centering
\caption{Continuation of Table\,\ref{tab:Log1}}
\resizebox{.47\textwidth}{!}{\begin{tabular}{clcccccc}
\hline  \hline
Index & Object & $\log T_{\rm eff}$ & $\log L$ & $X_{\rm H}$ & Age & $p_\text{N-rich}$ & References \\ 
 &  & kK & $[L_\odot]$ &  & Myr &  &  \\ 
\hline
53 & H40 & 47500 & 5.93 & 0.74 & 1.44 &  0.13 & 1;2\\ 
54 & VFTS 180 & 42300 & 5.92 & 0.45 &  - &  0.00 & 3;4\\ 
55 & VFTS 1018 & 42300 & 5.92 & 0.45 &  - &  0.00 & 3;4\\ 
56 & BAT99 79 & 45000 & 5.90 & 0.20 &  - &  1.00 & 9\\ 
57 & BAT99 99 & 45000 & 5.90 & 0.20 &  - &  1.00 & 9\\ 
58 & VFTS 169 & 47700 & 5.89 & 0.75 & 1.40 &  0.00 & 3;4\\ 
59 & VFTS 333-P & 33800 & 5.88 & 0.75 &  - &  0.00 & 3;4;13\\ 
60 & H58 & 47500 & 5.87 & 0.74 & 0.90 &  0.14 & 1;2\\ 
61 & VFTS 608 & 42300 & 5.86 & 0.75 &  - &  1.00 & 3;4\\ 
62 & VFTS 566 & 44900 & 5.85 & 0.75 & 1.70 &  1.00 & 3;4\\ 
63 & H50 & 47000 & 5.85 & 0.74 & 1.10 &  0.04 & 1;2\\ 
64 & H64 & 46000 & 5.85 & 0.74 & 1.44 &  0.28 & 1;2\\ 
65 & VFTS 216 & 44900 & 5.84 & 0.75 & 1.90 &  0.00 & 3;4\\ 
66 & H35 & 47500 & 5.82 & 0.74 & 1.10 &  0.00 & 1;2\\ 
67 & H45 & 41500 & 5.80 & 0.74 & 2.06 &  0.26 & 1;2\\ 
68 & BAT99 77 & 45000 & 5.79 & 0.70 &  - &  1.00 & 10\\ 
69 & BAT99 89 & 50000 & 5.78 & 0.20 &  - &  1.00 & 9\\ 
70 & Mk 33Na-S & 45800 & 5.78 & 0.74 & 1.60 &  0.00 & 12\\ 
71 & H55 & 47500 & 5.77 & 0.74 & 1.10 &  0.07 & 1;2\\ 
72 & H30 & 40000 & 5.76 & 0.74 & 2.22 &  1.00 & 1;2\\ 
73 & H49 & 44000 & 5.76 & 0.74 & 1.80 &  0.25 & 1;2\\ 
74 & VFTS 518 & 44900 & 5.75 & 0.75 & 0.70 &  1.00 & 3;4\\ 
75 & VFTS 532-P & 44900 & 5.74 & 0.75 &  - &  0.00 & 3;4;13\\ 
76 & H65 & 42000 & 5.73 & 0.74 & 2.18 &  0.33 & 1;2\\ 
77 & VFTS 664 & 39900 & 5.72 & 0.75 & 3.50 &  1.00 & 3;4\\ 
78 & VFTS 669 & 37700 & 5.71 & 0.68 &  - &  1.00 & 3;4\\ 
79 & H70 & 45500 & 5.71 & 0.74 & 1.32 &  0.21 & 1;2\\ 
80 & VFTS 422 & 39900 & 5.70 & 0.75 &  - &  0.50 & 3;4\\ 
81 & H52 & 45500 & 5.70 & 0.74 & 1.40 &  0.50 & 1;2\\ 
82 & H68 & 42000 & 5.68 & 0.74 & 2.16 &  0.84 & 1;2\\ 
83 & BAT99 76 & 35000 & 5.66 & 0.20 &  - &  1.00 & 9\\ 
84 & H66 & 47500 & 5.66 & 0.74 & 0.78 &  0.21 & 1;2\\ 
85 & VFTS 755 & 47700 & 5.64 & 0.15 & 1.70 &  0.00 & 3;4\\ 
86 & VFTS 455 & 42300 & 5.63 & 0.75 &  - &  0.50 & 3;4\\ 
87 & H62 & 45000 & 5.63 & 0.74 & 1.50 &  0.26 & 1;2\\ 
88 & VFTS 063-P & 44300 & 5.62 & 0.74 & 1.10 &  0.04 & 5\\ 
89 & VFTS 626 & 39900 & 5.60 & 0.60 & 2.60 &  1.00 & 3;4\\ 
90 & BAT99 120 & 32000 & 5.58 & 0.30 &  - &  1.00 & 9\\ 
91 & VFTS 797 & 44900 & 5.57 & 0.23 & 1.80 &  0.00 & 3;4\\ 
92 & VFTS 217-P & 45000 & 5.57 & 0.74 & 1.30 &  0.18 & 5\\ 
93 & VFTS 404-P & 47000 & 5.55 & 0.75 &  - &  0.00 & 3;4;13\\ 
94 & H78 & 46000 & 5.52 & 0.74 & 1.52 &  0.06 & 1;2\\ 
95 & VFTS 094-P & 41900 & 5.51 & 0.74 & 2.30 &  0.05 & 5\\ 
96 & BAT99 81 & 47000 & 5.48 & 0.40 &  - &  1.00 & 9\\ 
97 & H69 & 41000 & 5.47 & 0.74 & 2.36 &  0.29 & 1;2\\ 
98 & H71 & 45000 & 5.47 & 0.74 & 1.68 &  0.50 & 1;2\\ 
99 & H75 & 46000 & 5.47 & 0.74 & 0.06 &  0.31 & 1;2\\ 
100 & VFTS 450-S & 28300 & 5.46 & 0.74 & 7.30 &  0.90 & 5\\ 
101 & VFTS 171-P & 34250 & 5.43 & 0.75 &  - &  0.50 & 3;4;13\\ 
102 & BAT99 91 & 50000 & 5.42 & 0.20 &  - &  1.00 & 9\\ 
103 & H86 & 45500 & 5.38 & 0.74 & 0.60 &  0.15 & 1;2\\ 
104 & H94 & 44500 & 5.38 & 0.74 & 1.16 &  0.13 & 1;2\\ 
105 & VFTS 217-S & 41800 & 5.38 & 0.74 & 1.90 &  0.06 & 5\\ 
106 & H90 & 42000 & 5.35 & 0.74 & 1.94 &  0.28 & 1;2\\ 
107 & VFTS 538-S & 32000 & 5.31 & 0.74 & 4.90 &  0.96 & 5\\ 
108 & VFTS 450-P & 33800 & 5.30 & 0.74 & 4.60 &  0.13 & 5\\ 
109 & VFTS 500-P & 40700 & 5.29 & 0.74 & 2.30 &  0.05 & 5\\ 
110 & H92 & 40000 & 5.24 & 0.74 & 2.36 &  0.50 & 1;2\\ 
111 & VFTS 176-P & 38300 & 5.24 & 0.74 & 3.10 &  0.30 & 5\\ 
112 & VFTS 197-S & 33600 & 5.24 & 0.74 & 4.80 &  0.48 & 5\\ 
113 & VFTS 500-S & 39400 & 5.21 & 0.74 & 2.50 &  0.04 & 5\\ 
114 & VFTS 094-S & 40100 & 5.19 & 0.74 & 2.20 &  0.05 & 5\\ 
115 & VFTS 197-P & 33500 & 5.18 & 0.74 & 4.50 &  0.30 & 5\\ 
116 & H143 & 40000 & 5.16 & 0.74 & 2.60 &  0.26 & 1;2\\ 
117 & VFTS 063-S & 39300 & 5.13 & 0.74 & 2.10 &  0.05 & 5\\ 
118 & H80 & 35500 & 5.12 & 0.74 & 4.18 &  0.42 & 1;2\\ 
119 & VFTS 352-P & 41600 & 5.10 & 0.74 & 6.60 &  0.05 & 5\\ 
120 & VFTS 352-S & 40600 & 5.05 & 0.74 & 5.70 &  0.05 & 5\\ 
121 & VFTS 64-P & 33300 & 5.07 & 0.75 &  - &  1.00 & 13\\ 
122 & VFTS 174-P & 35900 & 5.00 & 0.74 & 3.70 &  0.04 & 5\\ 
\hline
\end{tabular}}
\label{tab:Log2}
\end{table}

\section{Nitrogen enriched probability at a fixed luminosity} \label{app:Nrich_frac}
Based on our full sample we aim to constrain the probability $P_\text{N-rich}(\log L/L_\odot)$ that a star with luminosity $\log L/L_\odot$ is nitrogen rich in the 30 Doradus region. For this purpose we model $P_\text{N-rich}(\log L/L_\odot)$ in terms of its value at fixed luminosities $\log (L_j/L_\odot)$, interpolating linearly in $\log (L/L_\odot)$ to obtain the probability for intermediate values of the luminosity:
\begin{align}
    P_\text{N-rich}(\log L/L_\odot) = P_\text{N-rich,j}\qquad\qquad\qquad\qquad\qquad\qquad\qquad \nonumber\\
    + \frac{P_\text{N-rich,j+1} -P_\text{N-rich,j}}{\log (L_{j+1}/L_\odot)-\log (L_j/L_\odot)}\left(\log (L/L_\odot)-\log (L_j/L_\odot)\right),
\end{align}
where $L_j \leq L < L_{j+1}$ and
\begin{align}
    P_\text{N-rich,j} = P_\text{N-rich}(\log L_j/L_\odot).
\end{align}
The objective is then to determine the values of $P_\text{N-rich,j}$ given the data and a specific choice of nodes $L_j$. In practice we use 21 value for $\log L_j/L_\odot$ separated by 0.1 dex between $\log L/L_\odot=5.0$ and $7.0$.

We constrain the values of $P_\text{N-rich,j}$ using a Bayesian model, where we aim to determine the posterior probability
\begin{align}
    P(\vec{\theta})\propto\mathcal{L}(\vec{d}|\vec{\theta})P(\vec{\theta}),
\end{align}
with $\vec{\theta}$ containing the model parameters $P_\text{N-rich,j}$, $P(\vec{\theta})$ representing the prior, $\vec{d}$ representing the observed stars in our sample, and $\mathcal{L}(\vec{d}|\vec{\theta})$ the likelihood.We consider a flat prior between zero and one for each of the $P_\text{N-rich,j}$. Additionally, since the exact luminosity $\log L_i/L_\odot$ of each star in our sample is not known, we also take each individual luminosity of each star to be part of the model parameters with a prior approximated as a split normal distribution based on the reported luminosity errors\footnote{In cases where no luminosity errors are available we assume a nominal error of $\pm 0.05$\,dex.}. The data $\vec{d}$ contains the individual probability of each star being nitrogen rich $p_\text{N-rich, i}$. For each star $i$ we consider an individual likelihood
\begin{align}
    \mathcal{L}_i = p_\text{N-rich, i}\times P_\text{N-rich}(\log L_i/L_\odot)\qquad\qquad\qquad\qquad \nonumber\\
    + (1-p_\text{N-rich, i})\times(1-P_\text{N-rich}(\log L_i/L_\odot)),\label{equ:L1}
\end{align}
with the total likelihood being given by the product of the likelihood of each star.
\begin{align}
    \mathcal{L}(\vec{d}|\vec{\theta})=\prod_i \mathcal{L}_i.\label{equ:L2}
\end{align}

To sample posteriors in this work we make use of the \texttt{Turing.jl} julia package \citep{Hong+2018, Fjelde+2025} together with the NUTS algorithm \citep{HoffmanGelman2011}. All samples produced in this work, as well as simulations, data products and code to reproduce results and figures, is publicly available at {\url{https://doi.org/10.5281/zenodo.20110444}}.

\section{Constraining $\alpha_\text{ov}$ and $f_\text{bin.\,prod.}$}\label{app:physical_constraints}
For a specific set of single star simulations that sample the parameter space of initial mass, overshoot parameter $\alpha_\text{ov}$, wind physics and star formation history, we can directly determine the enrichment probability of single stars at a given luminosity, $P_\text{N-rich,single}(\log L/L_\odot,\alpha_\text{ov})$ through population synthesis. For this purpose we assume a Salpeter initial mass function ($dN/dm\propto m^{-2.35}$, \citealt{Salpeter1955}), and take the star formation history to be either constant or follow a functional form motivated by the results of \citet{Schneider+2018_SFR},
\begin{align}
    \frac{\text{SFR}(t)}{M_\odot\,\text{yr}^{-1}} = \begin{cases}
        0.03 + \frac{t}{2\,\text{Myr}}(0.11-0.03) & t < 2\,\text{Myr} \\
        0.11 & 2\,\text{Myr} \leq t < 4\,\text{Myr}\\
        0.11 + \frac{t-4\,\text{Myr}}{4\,\text{Myr}}(0.02-0.11) & 4\,\text{Myr} \leq t < 8\,\text{Myr}\\
        0.02 & 8\,\text{Myr} \leq t,
    \end{cases}
\end{align}
where $t$ represents the amount of time before the present. This follows the trend of the full sample shown by \citet{Schneider+2018_SFR}, where the star formation history had its peak $\sim 2$\,Myrs ago, was relatively constant until $\sim 4$\,Myrs ago and drops prior to that.

Once $P_\text{N-rich,single}(\log L/L_\odot,\alpha_\text{ov})$ is determined, we also need to consider the fraction of stars $f_\text{bin.\,prod.}$ that we take to correspond to binary products that contribute equally at all luminosities. We take the probability of a star being nitrogen rich at a given luminosity (including effectively single stars and binary products) as
\begin{align}
    P_\text{N-rich}(\log L/L_\odot) = f_\text{bin.\,prod} \qquad\qquad\qquad\qquad\qquad\nonumber\\
    + (1-f_\text{bin.\,prod})\times P_\text{N-rich,single}(\log L/L_\odot,\alpha_\text{ov}).
\end{align}
making use of this enrichment probability we can make use of the same method described in Appendix \ref{app:Nrich_frac}, defining a likelihood following Equations (\ref{equ:L1}) and (\ref{equ:L2}) and performing and MCMC where the model parameters $\vec{\theta}$ include the individual luminosity of each star in the sample together with $\alpha_\text{ov}$ and $f_\text{bin.\,prod.}$.

\section{MESA Simulation setup}
In this work we consider three different mass loss prescriptions, including the ones used by \citet{Brott+2011}, \citet{Sabhahit+2022} and \cite{Brands2022}. For each of these wind prescriptions we compute a set of single star, non-rotating models, covering values of the step overshooting parameter $\alpha_\text{ov}=0.1$ to $4.0$ in steps of $0.05$ and masses between $\log M/M_\odot=1$ and $2.5$ (corresponding to $10M_\odot$ and $\sim 300M_\odot$ respectively) in bins of 0.05\,dex. Models were computed using the \texttt{MESA} stellar evolution code \citep{Paxton2011,Paxton2013,Paxton2015,Paxton2018,Paxton2019,Jermyn2023}. The initial composition was based on the tailored mixture for the LMC described in \cite{Brott+2011}, and we make use of custom OPAL opacity tables \citep{IglesiasRogers1996} that match this element mixture. We make use of the \texttt{basic.net} nuclear network of \texttt{MESA} which includes $^1$H, $^3$He, $^4$He, $^{12}$C, $^{14}$N, $^{16}$O, $^{20}$Ne and $^{24}$Mg. This network is sufficient to capture energy production and the transformation of carbon and oxygen into nitrogen during the main sequence \citep{Paxton2011}. Convection is modeled using the mixing length theory of \citet{Bohm-Vitense1958} as described by \citet{CoxGiuli1968} with a mixing length parameter of $\alpha_\text{MLT}=1.5$. Convective boundaries are determined using the Ledoux criterion \citep{Ledoux1947} with step overshoot determined by the choice of $\alpha_\text{ov}$. We also include semiconvective \citep{langer+1983} and thermohaline \citep{Kippenhahn+1980} mixing with efficiency parameters of unity, although these processes and their uncertainties are not expected to impact the evolution of single main sequence stars. For the regime studied the equation of state is based on the \texttt{FreeEOS} \citep{Irwin2004}. In order to increase the resolution of the sampled parameter space of mass and overshoot, we make use of the method of \citet{Dotter2016} to interpolate simulations.

\section{Reanalysis with reduced sample}\label{section:reanalysis}

\begin{figure}
   \includegraphics[width=\columnwidth]{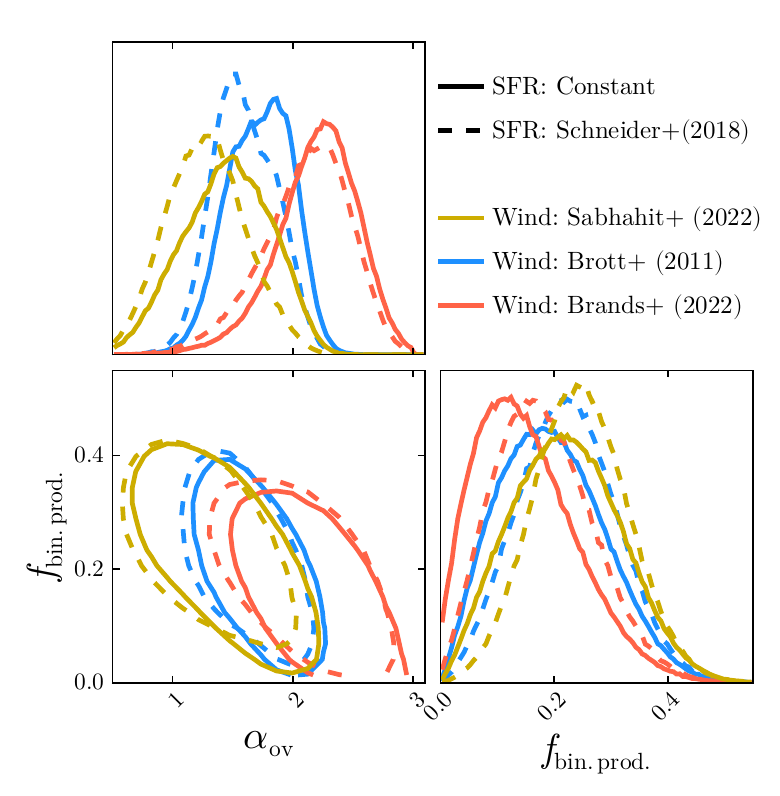}
   \caption{Same as Figure \ref{fig:triangle} but including only sources with surface hydrogen mass fractions $X>0.4$ to avoid possible contamination from classical Wolf-Rayet stars.}\label{fig:triangle_alt}
\end{figure}

In order to assess the possible impact of contamination from classical Wolf-Rayet stars, we repeated the inference of overshoot and binary fraction values including only sources with surface hydrogen mass fractions $X>0.4$. The results are shown in Figure \ref{fig:triangle_alt} and Table \ref{tab:inferred_reduced}. As expected all inferred values for overshoot are lower, but nevertheless peak well above unity indicating that our conclusion of enhanced mixing in very massive stars holds even when being very conservative about the sample.

\begingroup
\setlength{\tabcolsep}{8pt} 
\renewcommand{\arraystretch}{1.5} 
\begin{table}
\caption{Same as Table \ref{table:1} but including only sources with surface hydrogen mass fractions $X>0.4$ to avoid possible contamination from classical Wolf-Rayet stars.}              
\label{table:1alt}      
\centering                                      
\begin{tabular}{c | c c | c c}          
\hline\hline                        
 & \multicolumn{2}{|c|}{SFR: Constant} & \multicolumn{2}{|c}{SFR: Schneider} \\    
\hline
Winds & $\alpha_\text{ov}$ & $f_\text{bin.\,prod.}$ & $\alpha_\text{ov}$ & $f_\text{bin.\,prod.}$ \\    
\hline
Brott & $1.86^{+0.27}_{-0.56}$ & $0.18^{+0.16}_{-0.14}$ & $1.53^{+0.51}_{-0.33}$ &  $0.22^{+0.14}_{-0.15}$\\
Brands & $2.26^{+0.51}_{-0.59}$ & $0.12^{+0.16}_{-0.11}$ & $2.26^{+0.43}_{-0.74}$ & $0.16^{+0.16}_{-0.12}$ \\
Sabhahit & $1.50^{+0.56}_{-0.64}$ & $0.21^{+0.16}_{-0.15}$ & $1.34^{+0.48}_{-0.61}$ & $0.26^{+0.13}_{-0.15}$ \\
\hline                                             
\end{tabular}\label{tab:inferred_reduced}
\end{table}
\endgroup

\section{Additional mass loss variations}\label{app:mdot}
\begin{figure}
   \includegraphics[width=\columnwidth]{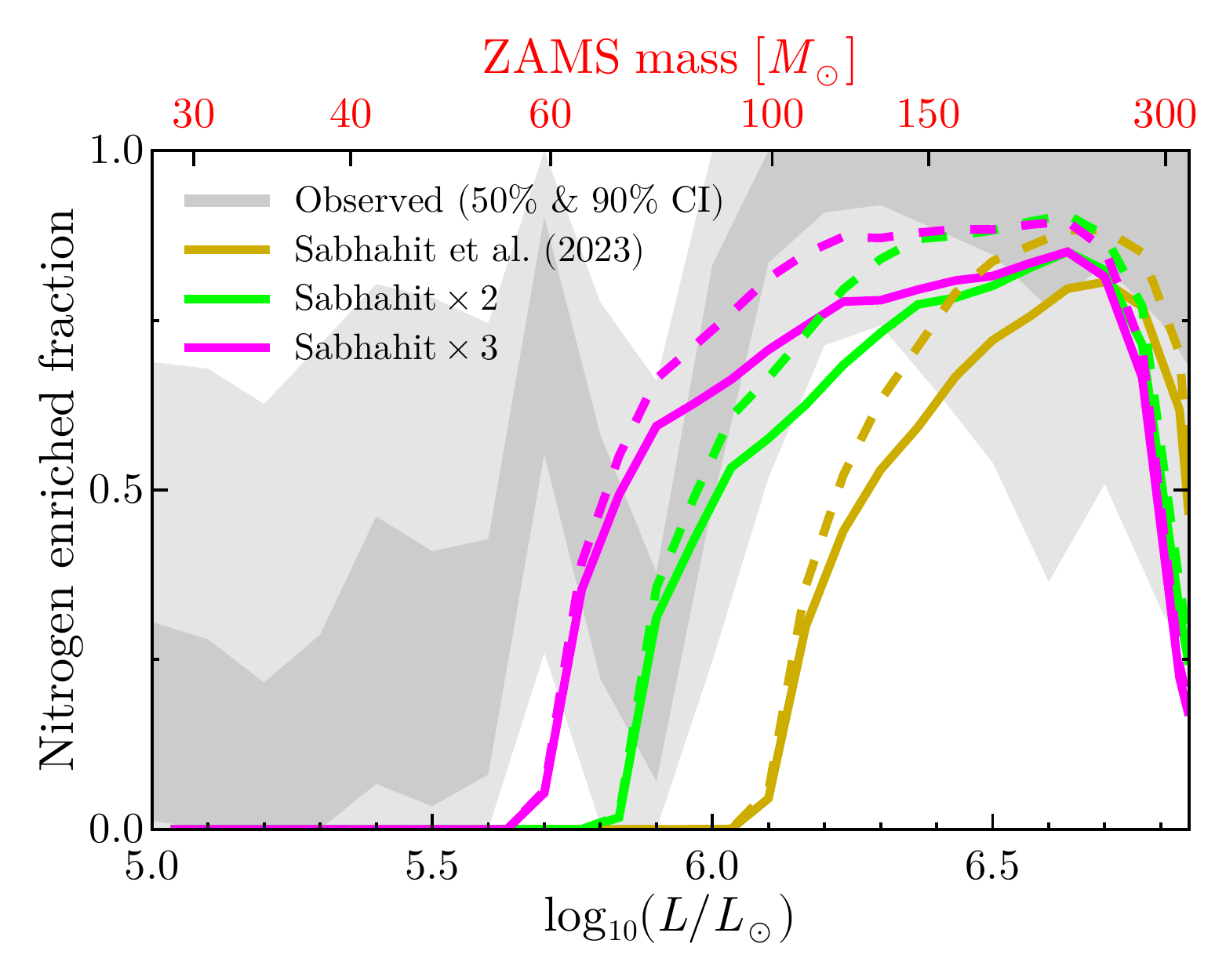}
   \caption{Same as Figure \ref{fig:enrichment_extra_wind} but showing results for the \citet{Sabhahit+2022} wind mass loss recipe increased by factors of two and three. Overshoot is taken to be $\alpha_\text{ov}=0.335$ following \citep{Brott+2011}.}\label{fig:enrichment_extra_wind}
\end{figure}

\begin{figure}
   \includegraphics[width=\columnwidth]{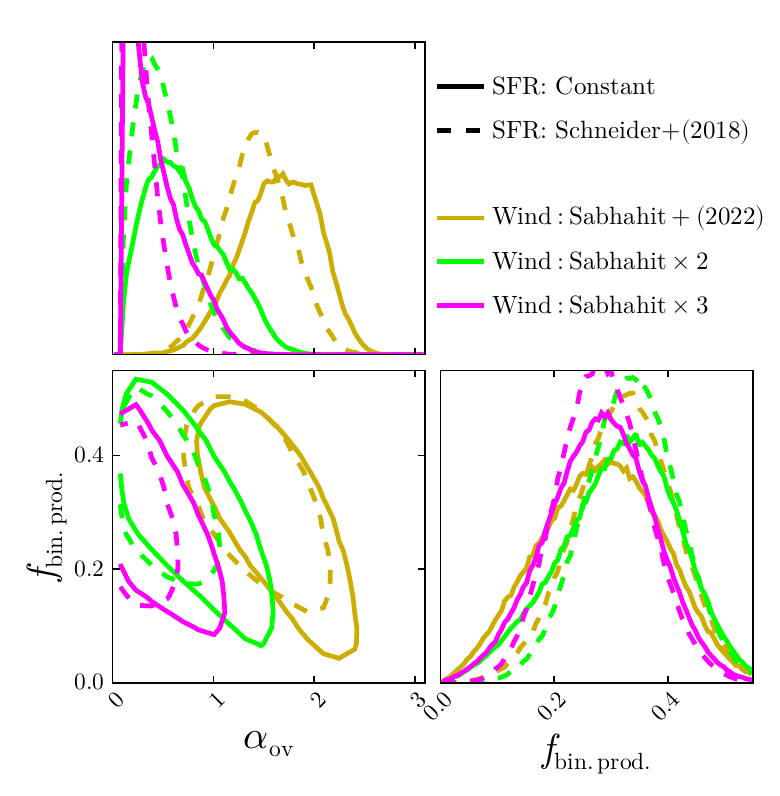}
   \caption{Same as Figure \ref{fig:triangle} but showing results for the \citet{Sabhahit+2022} wind mass loss recipe increased by factors of two and three.}\label{fig:triangle_extra_wind}
\end{figure}

As an additional test to see the impact of higher mass loss in our inference results, we consider the mass loss recipe of \citet{Sabhahit+2022} (which provides the lower inferred values of $\alpha_\text{ov}$) and repeat our analysis with a set of simulations that boost this recipe by factors of 2 and 3. As shown in Figure \ref{fig:enrichment_extra_wind} the stronger winds shift the luminosity at which simulations start showing enrichment downward, allowing a possible match with observations without the need for additional mixing. Figure \ref{fig:triangle_extra_wind} shows that in the extreme case where winds are taken to be 3 times higher than the recipe of \citet{Sabhahit+2022}, the inferred overshooting value peaks towards 0.1, the lowest value in our grid.

Despite these higher winds being formally able to explain the observed enrichment trends, below luminosities of $10^6L_\odot$ they rely on an increase on the mass loss recipe used by \citet{Sabhahit+2022} for optically thin winds, which is that of \citet{Vink+2001}. However, several constraints, both theoretical and observational \citep{Bjorklund+2023, Verhamme+2026} indicate that the mass loss recipe of \citet{Vink+2001} is already a significant overestimate. Observational constraints from \citet{Verhamme+2026} relate, however, to an overprediction in mass loss rates at temperatures below the bi-stability jump, which is expected to impact stars that have undergone significant evolution and is likely unrelated to the problem discussed in this manuscript which requires instead strong stellar winds early on in the evolution. The range of luminosities at which enhancement is observed is also that were stellar winds transition from optically thin to optically thick, so the mismatch observed with the unenhanced prescription of \citet{Sabhahit+2022} could indicate this transition happens at lower luminosities.

\begin{figure}
   \includegraphics[width=\columnwidth]{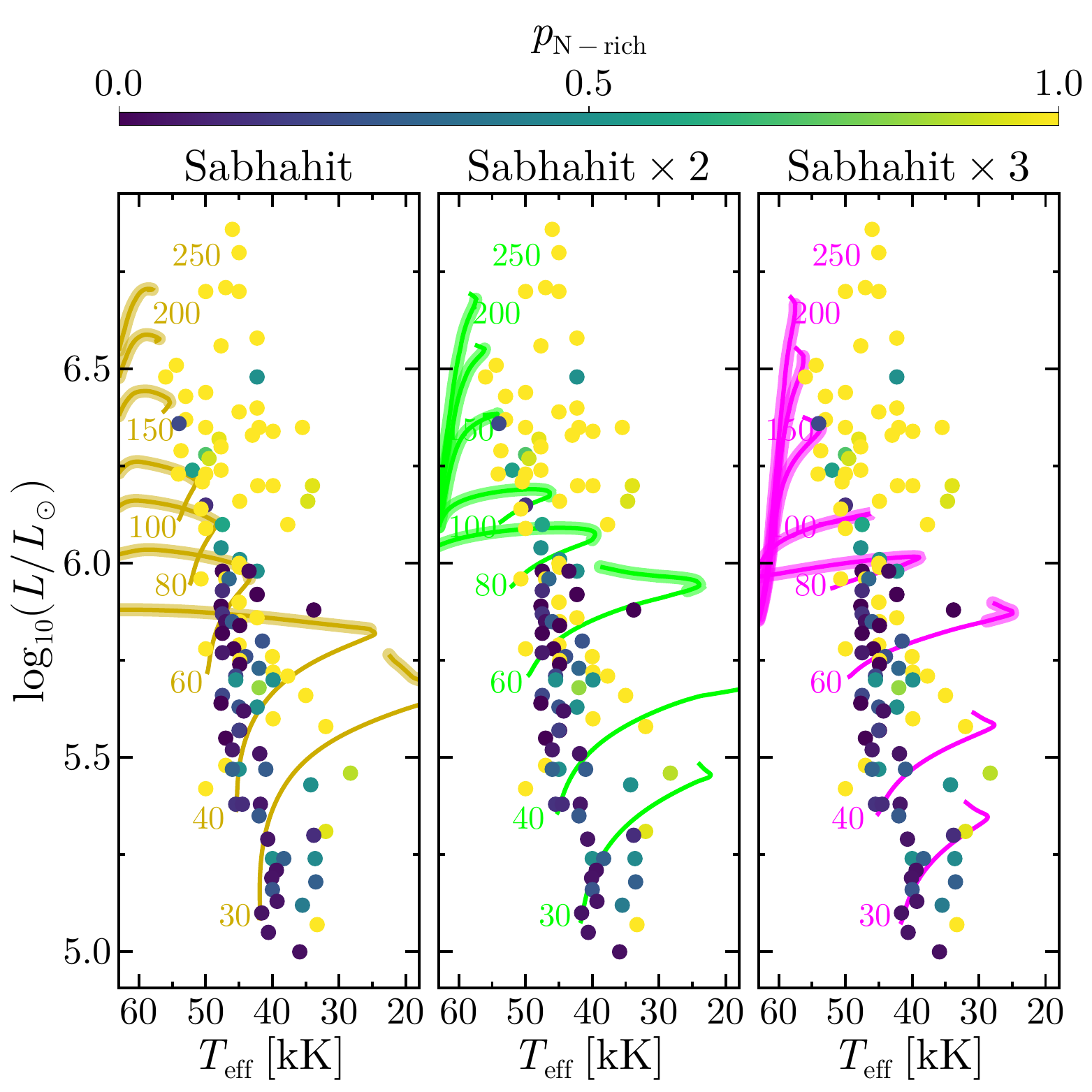}
   \caption{Same as Figure \ref{fig:HR_sample} but showing tracks for stars using the wind prescription of \citet{Sabhahit+2022} with different levels of enhancement. Overshoot chosen in each case corresponds to inferred values from our Bayesian analysis using the \citet{Schneider+2018_SFR} star formation history model ($\alpha_\text{ov}=$1.42, 0.356 and 0.1 respectively for each panel from left to right).}\label{fig:HR_sample_ov_extra_wind}
\end{figure}

Nevertheless, as illustrated by Figure \ref{fig:HR_sample_ov_extra_wind}, models adopting severely enhanced mass-loss rates fail to reproduce the observed HRD in a similar fashion to what is shown in Figure\,\ref{fig:HR_sample_ov}; to avoid this problem, one would need to invoke a sudden decrease of $\dot{M}$ along the evolution of the star, which lacks a theoretical or empirical motivation. The fact that these high $\dot{M}$ models contradict observations and theory drives us to the conclusion that enhanced mixing is a more likely candidate to explain the discrepancy in nitrogen enrichment reported on here.

\end{appendix}
\end{document}